\documentclass[12pt]{iopart}
\usepackage{epsfig}
\usepackage{amssymb}
\usepackage{amsfonts}
\usepackage{graphicx}

\begin{document}

\title[Chiral transition in a finite system and possible use of FSS in HICs]{Chiral transition in a finite system and possible use of finite size scaling in relativistic 
heavy ion collisions}
\author{L F Palhares$^{1,2}$, E S Fraga$^1$ and T Kodama$^1$}
\address{$^1$ Instituto de F\'\i sica, Universidade Federal do Rio de Janeiro,
Caixa Postal 68528, Rio de Janeiro, RJ 21941-972, Brazil} 
\address{$^2$ Institut de Physique Th\'eorique, CEA Saclay, 91191 Gif-sur-Yvette Cedex, France}
\ead{leticia@if.ufrj.br}
\ead{fraga@if.ufrj.br}
\ead{tkodama@if.ufrj.br}

\begin{abstract}
\noindent
We investigate finite-size effects on the chiral phase diagram of strong interactions within the linear 
sigma model coupled to quarks. We estimate the modification of the pseudocritical transition line 
and isentropic trajectories for sizes that are representative of the systems created in relativistic 
heavy ion collision experiments. The corrections are clearly non-negligible and might significantly 
affect signatures of the chiral critical endpoint based on estimates for an infinite system. We 
argue that a finite-size scaling analysis should be tested in the process of data analysis of the Beam 
Energy Scan program at RHIC and in future experiments at FAIR-GSI, through the use of full scaling 
plots and a chi-squared method as tools for searching the critical endpoint.
\end{abstract}


\submitto{\JPG}

\maketitle


\section{Introduction}

\noindent Although the process of phase conversion of hot hadronic matter
that presumably happens in ultra-relativistic heavy-ion collision (HIC)
experiments is often compared to the cosmological quark-hadron transition in
the early universe, the relevant space-time scales differ by almost twenty
orders of magnitude. In particular, while one would need temperatures $%
T\sim10^{8}~$GeV to be subject to appreciable finite-size effects in the
case of the early universe, which of course spares the primordial
quark-hadron transition from these effects, this is not the case for the
quark-gluon plasma formed in HICs \cite{Spieles:1997ab,Fraga:2003mu}.
Therefore, in the context of HICs, descriptions of an equilibrium
quark-gluon plasma in the thermodynamic limit are, in principle, missing
non-negligible finite-size corrections. These effects could be crucial for
the hydrodynamic description which is used to infer transport coefficients
and initial conditions \cite{Kodama:2006ga}.

As is well known (from the statistical mechanics standpoint), there is no
real phase transition in a system of finite size, and no criticality can be
defined. Singularities that would play the role of strong lighthouses in the
search for criticality are rounded and become smooth. Actually, the
originally critical quantities become \textit{pseudocritical}, with some
unavoidable ambiguity in their definition, and are shifted in the $T-\mu$
plane according to volume variations.

This is the actual scenario in any experiment in HICs. In addition, the time
scale involved in the collision process is also not large compared to the
microscopic time scale to achieve equilibrium. Therefore, thermodynamic
quantities directly inferred from data assuming an infinite system in
equilibrium should \textit{not} correspond to its true and unique value in
the thermodynamic limit. Moreover, even neglecting dynamical non-equilibrium
effects, all signatures of the second-order critical endpoint based on
non-monotonic behaviour \cite{Stephanov:1998dy,Stephanov:2008qz} (or sign
modifications \cite{Asakawa:2009aj}) of particle correlation fluctuations
will probe a \textit{pseudocritical endpoint}, shifted from
the genuine critical endpoint by finite-size corrections and boundary
effects.

In this paper, we do not address non-equilibrium effects, but show, using a
well-defined effective model for the chiral transition, that finite-size
effects can be very large in the physical systems under study in heavy ion
collisions. Hence, we may expect that the modifications brought
about by the finite volume of the system can play a crucial role in the
search for the critical endpoint in HICs. In HIC data, a 
direct access of the proposed particle correlation fluctuation
signatures to (volume- and boundary-condition-dependent) pseudocritical
quantities is very much restricted. On the other hand, we point out
that, since the experimental data can be arranged as a set of
systems of different sizes and incident energies, a finite size
scaling study might be helpful in extracting information directly related to
the thermodynamic limit of QCD, especially concerning its critical endpoint.

First, we investigate finite-size effects on the chiral phase diagram of
strong interactions, presenting results for the modification of the
pseudocritical line in the temperature-chemical potential phase diagram, as
well as for isentropic trajectories. For this purpose, we employ the linear
sigma model coupled to quarks with two flavors, $N_{f}=2$ \cite{GellMann:1960np}. This effective theory has been widely used to describe
different aspects of the chiral transition, such as thermodynamic properties 
\cite{quarks-chiral,Scavenius:2000qd,Scavenius:2001bb,Taketani:2006zg} and
the nonequilibrium phase conversion process \cite{Fraga:2004hp}, as well as
combined to other models in order to include effects from confinement 
\cite{polyakov}. We focus on the chiral phase transition, avoiding the inclusion
of confinement ingredients (although the linear sigma model coupled to
quarks contains mesons and quarks that alternate dominance in the two
different phases), and show that the volume-dependence of the phase diagram
of the model in the regime of energy scales probed by current heavy ion
experiments can be large. Even though the output numbers in an effective
model should not be taken as accurate predictions, the quantitative question
that matters here, and that can be answered within this approach, is how big
the dislocations of the pseudocritical line and critical point, isentropic
trajectories, etc are \textit{relative} to the same phase diagram in the
thermodynamic limit. As mentioned above they turn out to be non-negligible.

Strictly speaking, only systems that are infinite in volume exhibit
\textquotedblleft true\textquotedblright\ spontaneous symmetry breaking 
\cite{weinberg}. Nevertheless, for finite systems of volume $V$ the barrier
penetration factor is already suppressed as $\exp (-aV)$, $a$ being a
positive constant determined by the microscopic features of each system. Of
course, if one waits long enough, the system will be \textquotedblleft
distributed\textquotedblright\ between the two minima. If this suppression
is very effective, the two minima are physically well separated and
one can consider them two almost different
phases of the system, even if there is no true phase transition.

Keeping this caveat in mind, our phenomenological analysis is motivated by
the fact that heavy ion experiments strongly suggest a new phase of
deconfined matter \cite{Adams:2005dq} and the system is obviously finite,
and by the enormous success of this methods in the description of (\textit{%
pseudo})phase transitions in small systems \cite{DL,Cardy:1996xt}. For
simplicity, we disregard inhomogeneity corrections. Even though we are aware
that these are relevant (see Refs. \cite{Taketani:2006zg,Nickel:2009wj}),
they should not modify the general picture we present or diminish the
significance of the (large) finite size effects we obtain in this simpler
description.

Second, one should recall that 
HIC data can be separated into slices of different impact parameter width 
with different incident energies, i.e. a realization of a set of
thermodynamical systems of different sizes and temperatures. Thus, we
propose the use of full scaling plots of cumulants of fluctuations of pions
and an associated chi-squared method as tools that might be helpful in
searching for the critical endpoint of QCD in future analyses of heavy ion
data which cover a wide region of the phase diagram, such as those from the
Beam Energy Scan program \cite{BES} at RHIC-BNL and FAIR-GSI. From this
point of view, 
HICs could be seen as the \textquotedblleft experimental
realization\textquotedblright\ of lattice simulations for systems of
different sizes. In this picture, signatures of the second-order critical
endpoint based on non-monotonic behaviour (or sign modifications) of
particle correlation fluctuations will actually probe a \textit{set of
pseudocritical endpoints}, a feature which could contribute to the
broadening of the experimental signal, helping to wash it out in the large
thermal background. As in Monte Carlo simulations, the peculiarities of the
finite system become relevant (technically, this corresponds to the need of
inclusion of irrelevant operators in the formalism). In that case, one has
either to use the knowledge of the peculiarities of the system or resort to
finite-size scaling (FSS) techniques to extract indirect information about
criticality.


\section{Finite-size corrections to the chiral phase diagram}

Finite-size effects in the study of phase transitions via lattice
simulations, seen as an inevitable drawback of the method, and the necessary
extrapolation to the thermodynamic limit have been thoroughly studied for
decades \cite{DL}, the ultimate answer to the problem being given by the
method of FSS 
\cite{Cardy:1996xt}. Systematic calculations of finite-volume corrections
can then be computed not only near the criticality of continuous
(second-order) transitions, but also for first-order phase transitions 
\cite{binder-landau}, so that the thermodynamic limit can be taken in the
calculation of properties of the phase diagram.

However, for natural systems that are truly small, one should study the
modifications caused by its finitude in the phase diagram before comparing
to experimental observables. This is the situation in the study of the
chiral transition in 
HICs. Contrasting to the large number of studies in the computation of the
chiral condensate and related quantities within chiral perturbation theory
(see e.g. \cite{Damgaard:2008zs} and references therein), this issue is
often overlooked in the case of the quark-gluon plasma. Among the
exceptions, there is a lattice estimate of finite-size effects in the
process of formation of the quark-gluon plasma \cite{Gopie:1998qn}, a few
studies within the Nambu--Jona-Lasinio model \cite{finite-NJL}, and an
analysis of color superconductivity in dense QCD in a finite volume \cite{Yamamoto:2009ey}, besides investigations of finite-size effects on the
dynamics\footnote{For simplicity, we completely ignore the dynamics of the phase conversion.
Nevertheless, it was shown in Ref. \cite{Fraga:2003mu} that finite-size
effects will play a relevant role in the case of 
HICs.} of the plasma \cite{Spieles:1997ab,Fraga:2003mu}. The effect of
finite volume on the pion mass and chiral symmetry restoration at finite
temperature and zero density was investigated by Braun \textit{et al.} using
the proper-time renormalization group \cite{Braun:2004yk}.

To describe the chiral phase structure of strong interactions at finite
temperature and baryon density, we adopt the linear sigma model 
\begin{equation}
\mathcal{L} = \overline{\psi}_{f} \left[ i\gamma^{\mu}\partial_{\mu} +
\mu\gamma^{0} - g\sigma\right] \psi_{f} + \frac{1}{2}\partial_{\mu}\sigma%
\partial^{\mu}\sigma- V(\sigma,\vec{\pi})\;,  \label{lagrangian}
\end{equation}
where $\mu$ is the quark chemical potential, and 
\begin{equation}
V(\sigma,\vec{\pi})=\frac{\lambda}{4}(\sigma^{2} - \mathit{v}%
^{2})^{2}-h\sigma  \label{bare_potential}
\end{equation}
is the self-interaction potential for the mesons, exhibiting both
spontaneous and explicit breaking of chiral symmetry. We dropped the pion
terms for simplicity\footnote{%
The pion directions play no major role in the process of phase conversion 
\cite{Scavenius:2001bb}, so we focus on the sigma direction. However, the
coupling of pions to the quark fields might be quantitatively important in
the computation of the fermionic determinant inhomogeneity corrections \cite{Taketani:2006zg}.}. The $N_{f}=2$ massive fermion fields $\psi_{f}$
represent the up and down constituent-quark fields $\psi=(u,d)$. The scalar
field $\sigma$ plays the role of an approximate order parameter for the
chiral transition, being an exact order parameter for massless quarks and
pions. Quarks constitute a thermalized fluid that provides a background in
which the long wavelength modes of the chiral condensate evolve. We
incorporate quark thermal fluctuations in the effective potential for 
$\sigma $, i.e. we integrate over quarks to one loop. The parameters of the
lagrangian are chosen such that the effective model reproduces correctly the
phenomenology of QCD at low energies and in the vacuum. So, we impose that
the chiral $SU_{L}(2) \otimes SU_{R}(2)$ symmetry is spontaneously broken in
the vacuum, and the expectation values of the condensates are given by 
$\langle\sigma\rangle=\mathit{f}_{\pi}$ and $\langle\vec{\pi}\rangle=0$,
where $\mathit{f}_{\pi}=93$~MeV is the pion decay constant. The explicit
symmetry breaking term is determined by the PCAC relation which gives 
$h=f_{\pi}m_{\pi }^{2}$, where $m_{\pi}\approx138~$MeV is the pion mass. This
yields $v^{2}=f^{2}_{\pi}-{m^{2}_{\pi}}/{\lambda}$. The value of $\lambda=
20 $ leads to a $\sigma$-mass, $m^{2}_{\sigma}=2\lambda
f^{2}_{\pi}+m^{2}_{\pi}$, equal to $600~$MeV. The choice $g=3.3$, which
yields a reasonable mass for the constituent quarks in the vacuum, $M_{q}= g
f_{\pi}$, leads to a crossover at finite temperature, $T$, and vanishing
chemical potential, so that the first-order transition line that starts at
vanishing temperature and nonzero chemical potential stops at a critical
endpoint.

In the case of a finite system of linear size $L$, the momentum integral
from the one-loop quark contribution to the effective potential is
substituted by a sum 
\begin{equation}
\frac{V_{q}}{T^{4}}=\frac{2N_{f}N_{c}}{(LT)^{3}}\sum_{\mathbf{k}}\left[ \log
\left( 1+e^{-(E_{\mathbf{k}}-\mu )/T}\right) +\log \left( 1+e^{-(E_{\mathbf{k%
}}+\mu )/T}\right) \right] \;,
\end{equation}
where $E_{\mathbf{k}}=\sqrt{\mathbf{k}^{2}+m_{eff}^{2}}$, and 
$m_{eff}=g|\sigma |$ is the effective mass of the 
quarks\footnote{In general, the effects of finite size are not fully accounted for by the
replacement of continuum integrals by discrete sums over momentum states. In
certain cases, the importance of inhomogeneities can be enhanced in a small
system, especially near its surface. For simplicity, we neglect these
effects here.}. It is clear that finite-size effects in a quantum field
theory are, in a way, very similar to thermal effects with $1/L$ playing the
role of the temperature \cite{ZinnJustin:2000dr}. However, in the latter
boundary conditions are determined by the spin-statistics theorem, whereas
in the former there is no clear guidance (see, e.g., discussions in Refs. 
\cite{finite-NJL,Elze:1986db}). Choosing periodic boundary conditions (PBC)
are useful to focus on bulk properties of the model undisturbed by surface
effects \cite{ZinnJustin:2000dr,Brezin:1981gm,Brezin:1985xx}, but one is
always free to choose anti-periodic boundary conditions (APC) or any other,
instead. Depending on the physical situation of interest, a boundary
condition may be more realistic than others. To illustrate the dependence of
the pseudocritical phase diagram on the boundary conditions, we show results
for PBC and APC \footnote{Lattice studies \cite{Bazavov:2007zz} of the pure-gauge SU(3) deconfining
transition in a box with wall-type boundary conditions show sensibly larger
effects as compared to the analogous case with PBC. We also do not
investigate the influence of shape variations, choosing a cubic system.}.
For PBC, e.g., the components of the momentum assume the discretized values 
$k_{i}=2\pi \ell _{i}/L$, $\ell _{i}$ being integers, and there is a zero
mode (absent for APC, for which $k_{i}=\pi (2\ell _{i}+1)/L$). At zero
temperature, this zero mode in the case with PBC will modify the classical
potential, generating size-dependent effective couplings and masses.

\begin{figure}
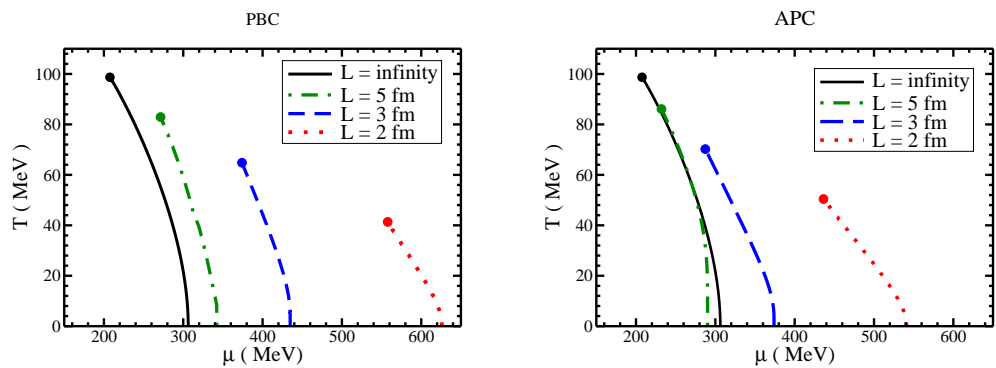

\center
\begin{minipage}[t]{64mm}
\includegraphics[width=6cm]{PhDiagram-PBC.eps}
\end{minipage}
\hspace{.4cm} 
\begin{minipage}[t]{70mm}
\includegraphics[width=6cm]{PhDiagram-APC.eps}
\end{minipage}
\noindent\caption{Pseudocritical transition lines and endpoints for different system
sizes within the linear $\protect\sigma$ model with periodic (left) and
antiperiodic (right) boundary conditions.}
\label{PhDiags}
\end{figure}

In what follows, we analyze systems of sizes \footnote{%
Our range of values for the linear size $L$ is motivated by the estimated
plasma size presumably formed in high-energy heavy ion collisions at RHIC 
\cite{Adams:2005dq}. The upper limit is essentially geometrical, provided by
the radius of the nuclei involved, whereas the lower limit is an estimate
for the smallest plasma observed.} between $10$ fm and $2$ fm which relate
to the typical linear dimensions involved in central and most peripheral
collisions of $Au$ or $Pb$ ions at RHIC and LHC, respectively. Figure \ref%
{PhDiags} displays the shift of the pseudocritical transition lines and
their respective endpoints as the size of the system is decreased. The
transition lines represent pseudo-first-order transitions, characterized by
a discontinuity in the approximate order parameter, the chiral condensate $%
\sigma$, and the production of latent heat through the process of phase
conversion. We find that those lines are displaced to the region of higher $%
\mu$ and shrinked by finite-size corrections. The former effect is sensibly
larger when PBC are considered, indicating that the presence of the spatial
zero mode tends to shift the transition region to the regime of larger
chemical potentials. Both boundary conditions reproduce the infinite-volume
limit for $L\gtrsim10$ fm. Figure \ref{CEP} shows the corresponding
displacement of the pseudocritical endpoint, comparing PBC and APC: both
coordinates of the critical point are significantly modified, and $\mu_{%
\mathrm{CEP}}$ is about $30\%$ larger for PBC. For $\mu=0$, the crossover
transition is also affected by finite-size corrections, increasing as the
system decreases, as shown in Figure \ref{Tcrossover}. Again, PBC generate
larger effects: up to $\sim80\%$ increase in the crossover transition
temperature at $\mu=0$ when $L=2$ fm.

\begin{figure}[tbh]
\vspace{0.5cm}
\center
\begin{minipage}[t]{66.5mm}
\includegraphics[width=6.3cm]{CEP.eps}
\caption{Displacement of the pseudocritical endpoint in the $T-\mu$ plane as the system size is
decreased for different boundary conditions.}
\label{CEP}
\end{minipage}
\hspace{.3cm} 
\begin{minipage}[t]{66.5mm}
\includegraphics[width=6.45cm]{Tcrossover.eps}
\label{Tcrossover}
\caption{Normalized crossover temperature at $\mu=0$ as a function of the inverse size $1/L$ for the cases with PBC and APC.}
\end{minipage}
\end{figure}

\begin{figure}[tbh]
\vspace{0.5cm}
\par
\center
\begin{minipage}[t]{64mm}
\includegraphics[width=6cm]{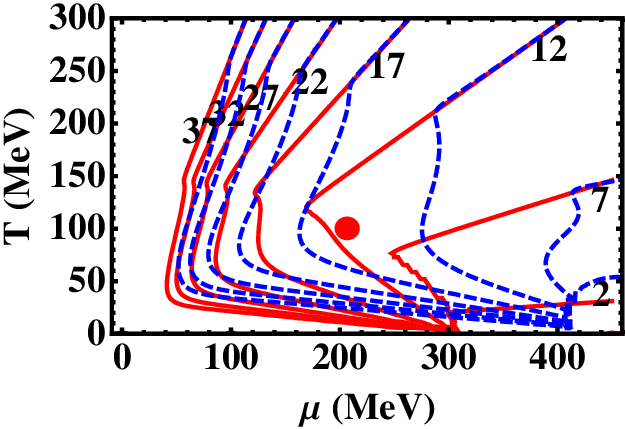}
\end{minipage}
\hspace{.5cm} 
\begin{minipage}[t]{64mm}
\includegraphics[width=6cm]{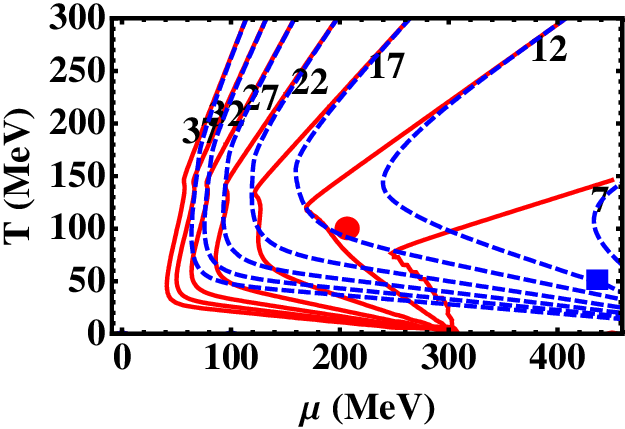}
\end{minipage}
\caption{Isentropic trajectories, labeled by the respective value of entropy
per baryon number, in the thermodynamic limit (solid, red lines) and for $%
L=2 $ fm with PBC (left) and APC (right). The red dot is the genuine
critical endpoint, while the square is pseudocritical one.}
\label{Isentropics}
\end{figure}

Results for the isentropic trajectories are shown in Figure \ref{Isentropics}%
, comparing the infinite-volume limit with the finite system with $L=2$ fm
in the cases with PBC and APC. For sufficiently high temperatures, the
isentropic lines in the thermodynamic limit are reproduced, while large
discrepancies are found around and below the transition region. The
remarkable variations at low temperatures are due to the shift of the
pseudocritical line and, for PBC, to the zero-mode-induced modifications of
the vacuum properties, especially the vacuum constituent quark mass.
Although there is no strong focusing effect around the critical endpoint (as
observed previously for this chiral model \cite{Scavenius:2000qd} and
similar ones \cite{Fukushima:2009dx}), it is clear that the density of
isentropic trajectories around the pseudocritical endpoint is sensibly
higher than in the thermodynamic limit.


\section{Possible application of FSS in HIC data analysis}

The close neighborhood of critical points exhibit strong and peculiar
scaling behavior that can be used as signals of its presence \cite%
{DL,Cardy:1996xt}. This fact has been recognized long ago as a possible
means of detecting experimentally the critical endpoint of QCD by using
event-by-event analysis in 
HIC experiments \cite{Stephanov:1998dy}. In fact, higher moments of
fluctuations of particle multiplicities diverge with increasing powers of
the correlation length \cite{Stephanov:2008qz}. As discussed in the
introduction and shown quantitatively for the chiral phase diagram in the
last section, finite-size corrections are relevant in the context of 
HICs, not only smoothening those singularities but also generating shifted
peaks and pseudocritical observables. Based on the analysis of the last
section, the amount of displacement of the pseudocritical transition
parameters that are probed at each experimental event should present a
significant centrality (size) dependence. Therefore, when considering data
from a centrality window, there is an average between different peaks which
should broaden even further the non-monotonic signal of the pseudocritical
endpoint. Furthermore, if the transition line is indeed strongly shrinked
and shifted to higher chemical potentials, the current experiments may not
be able to probe this regime of pseudocritical peaks and no non-monotonic
behaviour would be found in the data. Finally, if this is not the case, to
obtain, within this method, experimental information about the genuine QCD
critical endpoint, one would need to gather a set of pseudocritical endpoint
measurements for different system sizes and apply a procedure of
extrapolation to the thermodynamic limit, namely a 
FSS analysis.

Recently, it was proposed that a 
FSS analysis on top of the critical point would be a more efficient way of
searching \cite{Lizhu:2009md}, and, as mentioned previously, 
FSS was also used before in the context of the dynamics of a first-order
transition in QCD in a finite-volume quark-gluon plasma \cite%
{Spieles:1997ab,Fraga:2003mu}. The caveat of the aforementioned method for
searching experimentally the critical endpoint of QCD is the assumption that
one is essentially on top of the criticality. This condition will be hardly
met in 
HIC experiments, at least with enough statistics to stand out of the
background. Even if one is very close, or on top, of the critical point, a
system that is as small as the quark-gluon plasma created under these
conditions, where all relevant length scales are close \cite{Fraga:2003mu},
will be strongly affected by finite-size effects. Namely, all singularities
will be rounded and spread, something that can provide misleading artifact
``signals'' even in a 
FSS analysis when restricted to the assumption of being on top of the
critical point.

The alternative is simply making use of the full power of 
FSS, which is a method that is valid also when one is not so close to the
critical point\footnote{In a system with size $L$ and temperature $T$, the region where the scaling
becomes important is characterized by $L\sim \xi _{\infty }(T)=(\mathrm{%
length~scale})t^{-\nu }$.}, provided that we use full scaling plots. This
technique is predictive even for tiny systems in statistical mechanics (see,
e.g., Ref. \cite{sldq}), and, since it can be used in various different
regions of the phase diagram not far from the critical point, it can in
principle provide enough statistics for data analysis. In the
following, we suggest an application of this method to
search for the critical point of QCD (and possibly determining  its
universality class), assuming that we can build out of heavy ion data a set
of systems of different temperature and sizes via the incident
energies and the distribution of collisional centrality.

The 
FSS hypothesis \cite{DL,Cardy:1996xt,fisher}, that can be derived applying
the renormalization group to critical phenomena \cite{Brezin:1985xx,amit},
states that an observable $X$ in a finite system at temperature $T$ can be
written, in the neighborhood of criticality, in the form \cite{Brezin:1985xx}
\begin{equation}
X(t,g;\ell;L)=L^{\gamma_{x}/\nu} f_x(t L^{1/\nu}) \;,  \label{scaling}
\end{equation}
where $t=(T-T_{c})/T_{c}$ is the reduced temperature ($T_{c}$ being the
temperature associated with the critical point), $\gamma_{x}$ is the bulk
(dimension) exponent of $X$, $g$ the dimensionless coupling constant, $\ell$
a length to fix the renormalization procedure, and $\nu$ the critical
exponent of the correlation length that diverges as $\xi\sim t^{-\nu}$ at
criticality\footnote{The reduced temperature $t$ is a dimensionless measure of the distance to the critical point when no other external parameter is considered. In the
presence of other external parameters, such as $\mu$, one should redefine
this distance accordingly. }.
The dimensionful function $f_x(y)$ is universal up to scale fixing,
adjusting the overall dimension of the right-hand side of Eq. (\ref{scaling}).
 The critical exponents are sensitive essentially to dimensionality and
internal symmetry, which will give rise to the different universality
classes \cite{DL,amit}.

The critical contribution to experimental observables such as moments of
fluctuations of particle multiplicities near the critical point \cite{Stephanov:1998dy,Stephanov:2008qz} will suffer sizable corrections from the
function $f_{x}(y\neq 0)$ that can be computed within a model. One should
notice that the observables satisfying the scaling relation (\ref{scaling})
are the ones directly related to the correlation function of the order
parameter of the transition, such as fluctuations of the multiplicity of
soft pions, as discussed in Ref. \cite{Stephanov:1998dy}. Therefore, a
careful investigation of how the evolution of the system in the hadronic
phase and the large background of thermal correlations might affect or even
hide the scaling behaviour of a given experimental observable is demanded.

In this sense, instead of looking for the collapse of points
representing different sizes of the system at $t=0$ \cite{Lizhu:2009md}, plotting observables such as cumulants of
fluctuations of soft pions properly normalized\footnote{In this case the proper normalization will include not only the dimension
scaling associated with the exponent $\gamma _{x}$ (which contains in
principle a small anomalous contribution), but also the factors, such as
mesonic couplings, involved in the connection of the pion fluctuations to
the correlation functions of the order parameter of the chiral transition.}
as a function of the full scaling variable $tL^{1/\nu }$ can be an
efficient method for searching the critical endpoint of QCD and determining
its universality class. The critical temperature and chemical potential (the
coordinates of the critical endpoint) as well as the critical exponents
would be parameters in a fit (scaling plot), in the same manner as is done
in statistical mechanics simulations.

In the case of 
HICs one has to relate the scaling parameters with the experimental
observables, that is, one should find out how to arrange the data as
representing systems of different sizes and temperature.  The most natural
candidate related with the size parameter is the centrality, or the number
of participants, $N_{\rm{part}},$ of the collision (another 
possibility for assessing the approximate size of the system and its centrality dependence could be given by HBT interferometry data.). It is
interesting to note that when performing a scaling analysis, the scaling
variable needs to be defined only up to $L$- and $t$
-independent multiplicative factors. Therefore, the knowledge of the actual
size of the system during the phase conversion process is not needed, as
long as we have an observable that scales with it. We may write $L\sim
N_{part}^{q}$, where $q$ lies between $1/3$ and 
$1/2$, depending on the geometrical conditions of the collision. In
this paper, we assume $q=1/2.$

Describing the distance to the CEP in HICs is a more subtle issue,
because there are two important external parameters(
actually, there are also two different critical exponents $\nu _{T}$ and $%
\nu _{\mu }$, since the divergence of the correlation length will be
different in these two directions, as it happens for the Ising model with
temperature and external magnetic field.): the temperature $T$ and
the chemical potential $\mu $. Determining (or even defining, in
the case of a dynamics far from equilibrium) the direction in which the
system created in a HIC approches a given point in the phase diagram is an
extremely difficult task. Fortunately, FSS is a very general and flexible
technique and does not require this precise piece of information, but rather
a reasonable measure of the distance between the CEP and the external
parameters that characterize the data to be analyzed. As suggested by the
success of ideal hydrodynamic models, the expansion of the system may well
be considered adiabatic. Therefore, we expect that the point of the emission
of soft pions should be on this adiabatic line, characterized by the initial
energy density. In this vein, the reduction in the number of dimensions in
the parameter space seems very natural. Since the adiabat curve in$\left(
T-\mu \right) $ plane is naturally parameterized by the
center-of-mass energy $\sqrt{s}$ of the collision, one simple and
reasonable prescription is to define the distance up to the first order by $%
\delta s\equiv (\sqrt{s}-\sqrt{s_{c}})/\sqrt{s_{c}}$, where $\sqrt{%
s_{c}}$ is the center-of-mass energy for which the adiabat passes 
the genuine CEP. 

When one plots $XL^{-\gamma _{x}/\nu }$ as a function of,
e.g., $T$ (not a scaling plot), one finds scattered curves that
cross in the critical point. In a full scaling plot, though, that uses the
correct scaling variable, all these curves collapse, provided that the
system is not far from the critical point, 
as a consequence of scaling and universality. By finding the correct
critical exponents in the case of the QCD phase transition in 
HICs, one would determine its universality class. One can search for this
full scaling behavior in different cumulants of fluctuations and in
different parts of the phase diagram, increasing dramatically the 
statistics. Realistically, the thermal background of particle 
production could hide the
critical contribution inside the error bars and the process of finding the
correct value of the critical exponent could be difficult.

In practice we may, for example, consider the following optimization
procedure. Suppose a full set of scaling observables  $X$ (say,
cumulants of soft pion fluctuations) is obtained as a function of $\left( 
\sqrt{s},N_{\mathrm{part}}\right).$ Let us then introduce the
quantity defined by%
\begin{equation}
Q\left( \sqrt{s},\sqrt{N_{\mathrm{part}}},\nu \right) =X\left( \sqrt{s},N_{%
\mathrm{part}}\right) ~N_{\mathrm{part}}^{-\gamma _{x}/2\nu }.  \label{Qeq}
\end{equation}%
From Eq.(\ref{scaling}), this should be a function of the scaling
variable $y=N_{\mathrm{part}}^{\nu /2}~\delta s$ only. Therefore,
we may introduce the following function,%
\begin{equation}
\chi ^{2}\left( \nu ,\sqrt{s_{c}};y\right) =\sum_{N_{\mathrm{part}}^{\nu
/2}~\delta s\,=\,y}\omega \left( \sqrt{s}\,,\,\sqrt{N_{\mathrm{part}}}%
\right) \left( \frac{Q\left( \sqrt{s}\,,\,\sqrt{N_{\mathrm{part}}}\,,\,\nu
\right) }{Q\left( \sqrt{s}_{0}\,,\,\sqrt{N_{\mathrm{part},0}}\,,\,\nu
\right) }-1\right) ^{2}  \label{chi2}
\end{equation}%
where the sums are taken over all pairs of $\left( \sqrt{s},\sqrt{
N_{\mathrm{part}}}\right) $ keeping the scaling variable $y=y_{0}$
 constant, with %
\begin{equation}
N_{\mathrm{part},0}^{\nu /2}~\delta s_{0}\equiv y\,_{0},
\nonumber
\end{equation}%
and $\omega \left( \sqrt{s}\,,\,\sqrt{N_{\mathrm{part}}}\right) $
 is a suitable weight function associated with the errors for each
point. If all of $X$ values for different $\left( \sqrt{s},\sqrt{%
N_{\mathrm{part}}}\right) $ obey finite-size scaling, then the
above $\chi ^{2}$ should vanish at the correct value of $\nu $
 and $\sqrt{s_{c}}$ for any $y$. For energies at
which the observables attain values close to the critical point, then we
should have a very week $y$-dependence of $\nu $ and $%
\sqrt{s_{c}}$ and also the value of $\chi ^{2}$ at the
minimum, coming exclusively from the statistical errors. In practice, we plot $%
\chi ^{2}$ for a given $y$ (or various combinations of $%
\left( \sqrt{s}_{0},\sqrt{N_{\mathrm{part},0}}\right) $ and
localize the minimum in the parameter space of $\nu $ and $\sqrt{s_{c}}$.

Thus, by studying the energy dependence of the minimum of the above
quantity, we may identify the energy value which passes close to the
critical point with the free variational parameters being the critical
exponent $\nu $ and the critical center-of-mass energy $\sqrt{s_{c}}$ (and $%
\gamma_x$ in the general case).

Eq. (\ref{chi2}) was built in this paper to illustrate how one can define a
chi-squared method to analyze scaling plots in this context. One can
certainly build other functions of this sort, and we are sure that
experimental groups will come up with alternatives that are more adapted to
the subtleties of data analysis. Our point is the possible usefulness of the
power of full scaling plots (within finite-size scaling) to search for
criticality in QCD and heavy ion collisions. Near criticality, Eq. (\ref%
{scaling}) can be rigorously demonstrated within the renormalization group
framework \cite{amit}, and our definitions are thereby solid. Once the
appropriate variables for scaling are identified, and these are built from a
measure of the linear size of the system and of the dislocation of the
system from the critical point (proportionality being enough), the scheme is
well defined. We built these variables from $N_{part}$ and $\delta s$,
quantities that are physically meaningful and commonly used by
experimentalists in their analyses.


\section{Final remarks}

In this paper, we have discussed different aspects of finite-size effects
that render crucial features in various proposed signatures of the QCD
critical endpoint in 
HICs. On one hand, in addition to the usual smoothening and broadening of
singularities, the finiteness of the physical system under a phase
conversion process shifts pseudocritical observables associated with
pseudocritical parameters with respect to the genuine criticality, in the
thermodynamic limit. We have shown that those displacements in the case of
the chiral transition can be large for energy scales typically encountered
in current experiments, shrinking the $T$-$\mu$ domain where discontinuities
of the chiral condensate are found and pushing the pseudocritical lines
further into the high chemical potential region. These findings should
affect all signatures of the critical endpoint based on non-monotonic
behaviours or sign changes in particle correlation fluctuations, probably
contributing to the broadening of the signal within a not sufficiently small
centrality window. One should note, however, that the results were obtained
within the mean-field approximation, which is an obvious caveat regarding
the precise size of the effects.

Based on this considerations we proposed the use of the full power of the
FSS hypothesis as a pragmatic tool to detect and locate the genuine chiral
critical endpoint of QCD, and determine its corresponding universality
class, using HICs experiments. We also described a chi-squared method that
can be applied to data in a systematic manner to investigate the presence of
the scaling phenomenom in the Beam Energy Scan project at RHIC and in future
observations at FAIR. Of course, since the pragmatic trial scaling relation
proposed can not be derived from first principles, only its direct
application to the analysis of actual data will be the final test of the
method, as with any pragmatic procedure of this nature. A detailed study of
the implementation of the FSS analysis to HIC data is currently under way 
\cite{futureP}.

Finally, to address the question of the viability of these different
proposed signatures of the QCD critical endpoint when immersed in a large
hadronic thermal background, as the one expected in the case of HICs, an
interesting test to make would be the comparison of Monte Carlo simulations
of thermal ensembles with and without a small fraction of critically
correlated soft pions.

\ack

We are grateful to M. Chernodub, L. Moriconi, S.L.E. de Queiroz and P.
Sorensen for discussions. This work was partially supported by CAPES, CNPq,
FAPERJ and FUJB/UFRJ.


\end{document}